# Determination of new national highpoints of five African and Asian countries – Saudi Arabia, Uzbekistan, Gambia, Guinea-Bissau, and Togo

Eric Gilbertson[1], Matthew Gilbertson

1. Department of Mechanical Engineering, Seattle University, Seattle, WA, USA

**Abstract:**

Not all nations on earth have previously been surveyed accurately enough to know for certain which peak is the national highpoint, the highest peak in the country. Knowledge of these peaks is important for understanding the physical geography of these countries in terms of natural resource availability, watershed management, and tourism potential**.** For this study, ground surveys were conducted between 2018-2025 with modern professional surveying equipment, including differential GPS units and Abney levels, to accurately determine the national highpoints in five African and Asian countries where uncertainty existed. New national highpoints were determined for Saudi Arabia (Jabal Ferwa), Uzbekistan (Alpomish), Gambia (Sare Firasu Hill), Guinea-Bissau (Mt Ronde), and Togo (Mt Atilakoutse). Elevations were measured with sub-meter vertical accuracy for candidate peaks in Saudi Arabia, Gambia, Guinea-Bissau, and Togo. Relative elevations were measured between contender peaks in Uzbekistan with sufficient accuracy to determine the highpoint.

## 1 Introduction:

The location of the highest peak in a country is a source of geographic significance and national pride. Most, but not all, countries have been surveyed accurately enough to determine the location and elevation of the highest peak. Knowing this information is important for economic, cultural, geographic, and climatological reasons. For instance, on average 60,000 climbers and trekkers visit the Mount Everest region of Nepal each year, contributing millions of dollars to the local economy (SNP 2025). In Japan, over 300,000 people annually climb Mt Fuji (Fuji 2025). National highpoints often hold cultural significance, such as Mt Olympus in Greece, which was the mythological home of the gods and is now a UNESCO Biosphere Reserve. Knowing accurate information about high peaks in a country can also inform decisions on natural resource management, as these peaks are often in major watershed areas (Dennedy-Frank et al. 2024). Many climate models depend on knowing accurate mountain elevations (Mountain Research 2015).

Satellite-based elevation measurements exist for all countries on earth, but measurement errors are still high enough to be significant in determining which of a collection of potential national highpoints is the highest. In February 2000, the satellite-based Shuttle Range Topography Mission (SRTM) collected elevation data at discrete points between 56 degrees south and 60 degrees north latitude at 1 arcsecond spacing (approximately 30m) with reported vertical accuracy +/-16m (Smith 2023). Elevations of locations between measured points have been approximated by different Digital Elevation Models (DEMs). However, the error bounds of elevations of locations between measured points is unknown and can potentially be higher than +/-16m, especially for sharp peaks (Sandip 2013).

Airplane- or drone-based LiDAR measurements can achieve vertical errors less than +/-0.3m (ASPRS 2004) but have only been conducted in a limited number of locations around the world. Traditional theodolite-based ground surveys have been conducted in most countries; however error bounds are often not reported in elevation measurements or on topographic maps. Sometimes different theodolite-based ground surveys resulted in different elevations for the same peak, adding to uncertainty. In other instances, some but not all, of the highest peaks in a country were surveyed, resulting in unsurveyed peaks with approximate but unknown elevations on topographic maps.

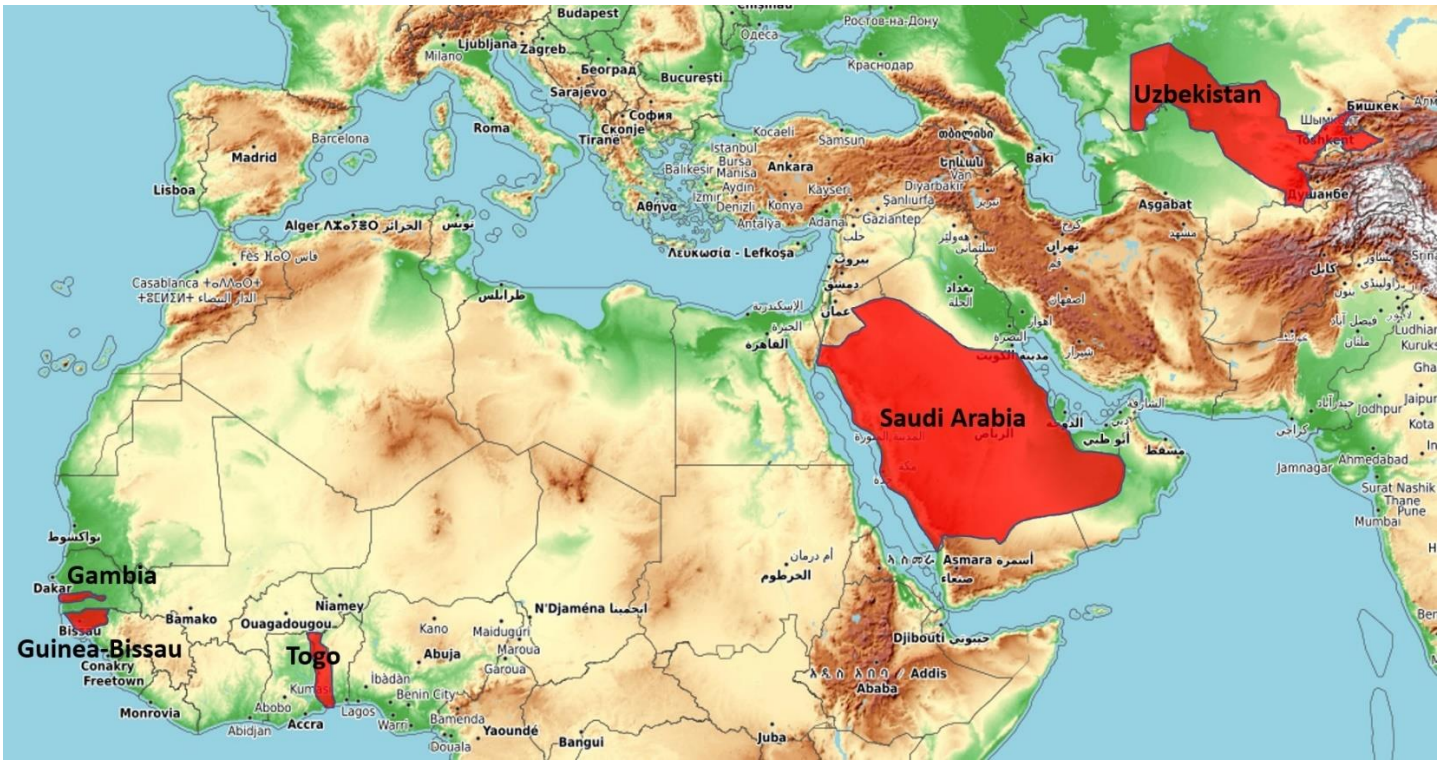

*Figure 1: Locations of the five countries where ground surveys were conducted -- Saudi Arabia, Uzbekistan, Gambia, Guinea-Bissau, and Togo. Basemap from OpenTopography (OpenTopography 2025).*

Differential GPS units have been used since the 1990s to measure elevations with vertical errors less than +/-0.03m (Schrock 2011). This measurement method generally requires a unit to be mounted in a fixed location for at least one hour. Measurements with these units have been used to determine elevations of national highpoints of countries including France/Italy (Vincent et al. 2007, Berthier et al. 2023), China/Nepal (Junyong 2005), Pakistan (Lehmuller 2015), Peru (Huascaran 2017), USA (Wagner 2015), the Dominican Republic (Geospatial 2021), and Lithuania (Krupick 2007).

There is only one previous instance where ambiguity in the location of a national highpoint was resolved with modern surveying methods like differential GPS. In 2018 differential GPS measurements were used to discover that the national highpoint of Sweden had switched from the southern summit of Kebnekaise to the northern summit (Holmlund 2019). This switch occurred because the southern summit of Kebnekaise, an icecapped summit, melted down enough that the northern, rocky summit became the true country highpoint.

Prior to this study, there existed 14 countries in the world with ambiguous national highpoints, located in South America, Africa, Asia, and Oceania. These countries had not previously been surveyed accurately enough to definitively determine the locations and elevations of the national highpoints. The countries are Colombia, Botswana, Benin, Myanmar, Marshall Islands, Maldives, Zambia, Guyana, the Federated States of Micronesia, Saudi Arabia, Uzbekistan, Gambia, Guinea-Bissau, and Togo.

Between 2018 and 2025 the authors conducted ground surveys using differential GPS units and Abney levels to definitively determine the national highpoints in five of these countries: Saudi Arabia, Uzbekistan, Gambia, Guinea-Bissau, and Togo (Figure 1).

This paper describes the measurement methodology, analysis of results, and comparison to satellite-based measurements and previous ground surveys for each of these five countries. For each country, a new highpoint was identified that was different than the previously-accepted national highpoint.

Section 1 presents background information summarizing historical survey data for the candidate highpoints in the five countries. Section 2 describes the methods used to conduct surveys from 2018-2025, Section 3 presents the results, and Section 4 summarizes the findings.

The authors have additionally conducted surveys in Colombia in 2024, and that is the subject of a different paper. The peaks surveyed in Colombia have icecapped summits which are melting down, a different situation than the peaks in this study. The remaining eight countries represent potential future work.

## 1.1 Saudi Arabia

The two leading candidates for the highest point in the Kingdom of Saudi Arabia, based on previous ground surveys and satellite-based DEMs, are Jabal Sawda (location 18.266717N, 42.368264E) and Jabal Ferwa (location 17.928547N, 43.265528E), both of which are located in southern Saudi Arabia (Figure 2A).

According to Google Maps DEM, Jabal Sawda, previously considered the highest peak in the country, lies between the 2980m and 3000m height contours (Google Terrain 2018). There is one (and only one) other peak in Saudi Arabia, Jabal Ferwa, previously thought to be the second highest mountain, that also lies between the 2980m and 3000m contours on this DEM. A different DEM, Google Earth, measured each peak to be the same elevation, 2987m (Google Earth 2018). Both DEMs use the EGM96 geoid.

Several land-based topographic surveys have been conducted in Saudi Arabia that measured one or both peaks. The Soviet Military conducted a survey in 1978 that measured Jabal Sawda to have an elevation of 3032m and Jabal Ferwa to have an elevation of 3091m (Soviet E38-13 1978).

The US military conducted a survey in 1980 (US J06 1980) that measured Jabal Sawda at 3207m and Jabal Ferwa at 2941m. The US military conducted another survey in 2001 (US J06-C 2001) that measured Jabal Sawda at 3015m and Jabal Ferwa at 3020m. These were the only ground surveys that measured both peaks. A US Military 1957 ground survey (US NE38 1957) measured Jabal Sawda at 2910m, but did not measure Jabal Ferwa. These surveys used standard triangulation methods and error bounds were not given on measurements.

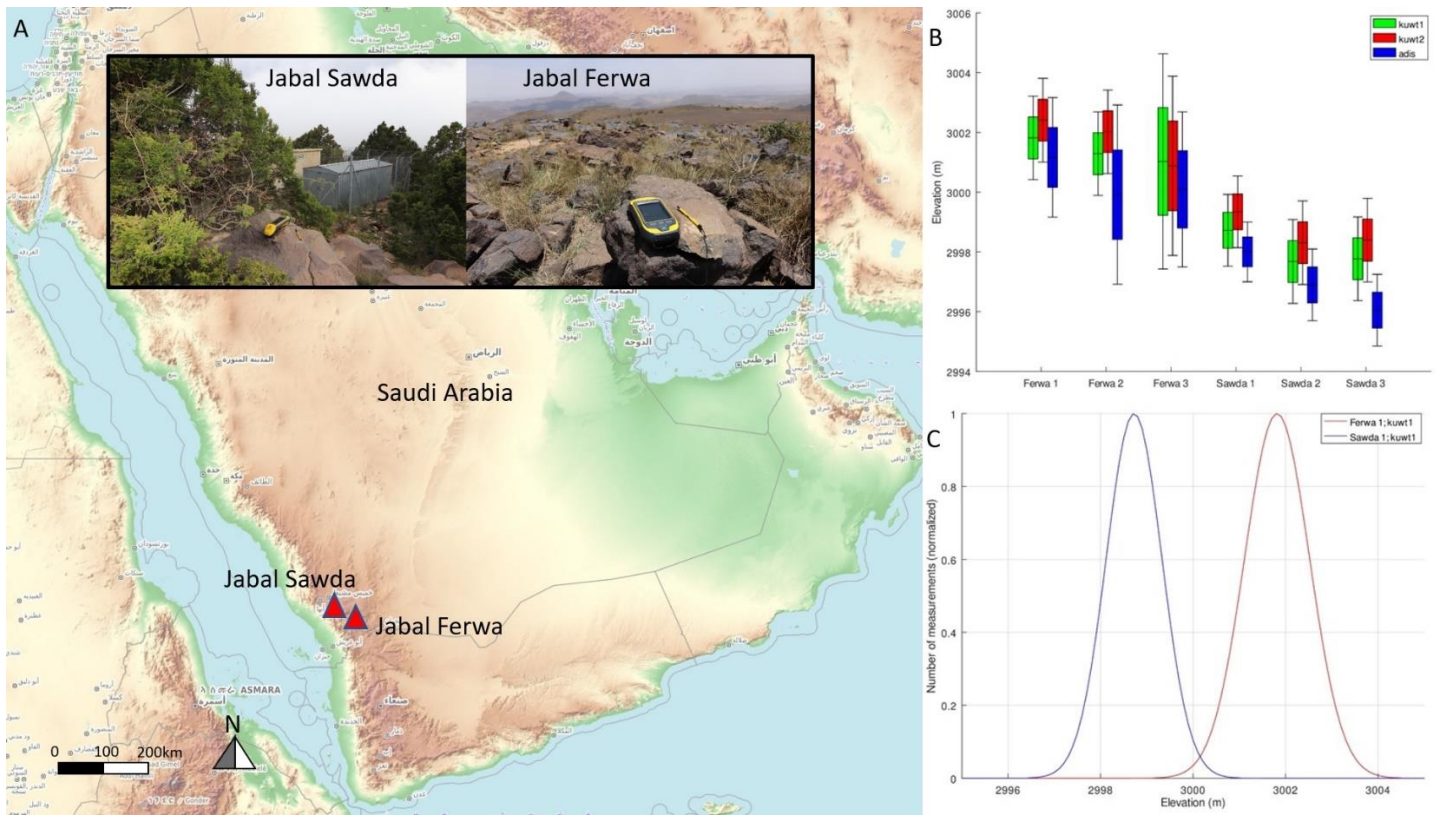

Figure 2. (A) The locations of Jabal Sawda and Jabal Ferwa, the two highest peaks in Saudi Arabia. Basemap from OpenTopography (OpenTopography 2025). (B) orthometric elevation for each peak for three different post-processing methods. Box height represents one sigma error and whisker represents two-sigma error. Center horizontal line is the mean elevation. Box color refers to post-

*processing method. (C) normalized histograms of the elevation measurements for Ferwa 1 and Sawda 1 waypoints, assuming a normal distribution.*

A summary of the measured elevations (in meters) of Jabal Sawda and Jabal Ferwa is presented in Table 1. The geoid model was not specified on triangulation-based survey maps.

*Table 1: Elevation measurements of Jabal Sawda and Jabal Ferwa.*

| Survey | | | Mountain | | |
|---|---|---|---|---|---|
| Name | Method | Year | Sawda | Ferwa | Error (m) |
| US Military | Triangulation | 1957 | 2910 | n/a | ? |
| Soviet Military | Triangulation | 1978 | 3032 | 3091 | ? |
| US Military | Triangulation | 1980 | 3207 | 2941 | ? |
| Google Terrain | Satellite, DEM | 2000 | 2980-3000 | 2980-3000 | +/-16 |
| Google Earth | Satellite, DEM | 2000 | 2987 | 2987 | +/-16 |
| US Military | Triangulation | 2001 | 3015 | 3020 | ? |

The prior survey data call into question whether Jabal Sawda is indeed the true highest point in Saudi Arabia. Previous ground surveys were inconclusive which peak was taller, with some surveys measuring Jabal Sawda taller and others measuring Jabal Ferwa taller, with no error bounds provided for measurements.

Satellite-based measurements found the peaks the same elevation within the error bounds of the measurements.

## 1.2 Uzbekistan

Before 2023, it was thought that Peak 4643 (location 38.948396N, 68.172312E, sometimes incorrectly called Khazret Sultan) was the highest peak in Uzbekistan (UNFCCC 1999). Peak 4643 is located on the border with Tajikistan in the Gissar Range (Figure 3A). One ground survey has been conducted in the region, in 1980, by the Soviet Military (Soviet J42-41 1980).

This survey produced a topographic map with the highest peak surveyed at 4643.3m. This peak was previously named "Peak of the 22$^{nd}$ Congress of the Communist Party" before Uzbekistan gained independence, but after independence this name was dropped. It is currently referred to in Uzbekistan by its surveyed elevation, Peak 4643.

Before 2023 Peak 4643 had just four documented ascents - the original survey team, plus ascents in 2005 (Fullen 2005), 2010, and 2018.

Satellite-based measurements using the Google Earth DEM (Google Earth 2023), however, showed a second peak in Uzbekistan of equal elevation within the error bounds of the measurements (Figure 3A). This peak has been referred to by mountaineers as Alpomish (location 38.891634N, 68.176775E).

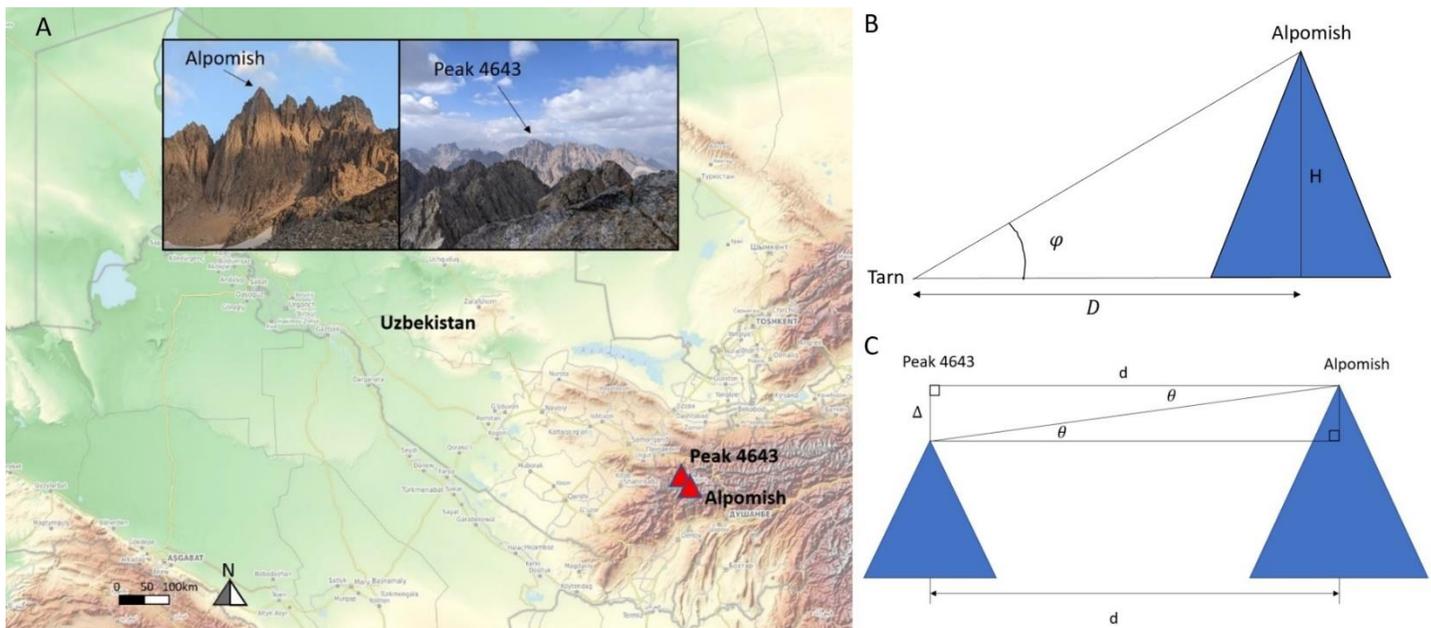

*Figure 3. (A) Map showing location of Alpomish and Peak 4643, the two highest peaks in Uzbekistan. Basemap from OpenTopography (OpenTopography 2025). (B) Measurement of Alpomish altitude from camp at tarn. (C) Angular inclination/declination measurements between Alpomish and Peak 4643.*

The Google Earth DEM estimated Alpomish 2m taller than Peak 4643 (Google Earth 2023), the Google Terrain DEM estimated both within the same 20m contour interval (Google Maps 2023), and the Opentopo DEM estimated Alpomish 19m taller than Peak 4643 (Opentopomap 2023) (Table 2). These elevations are all within the error bounds of DEMs, so the data is inconclusive for which peak is higher. Satellite-based DEMS are reported using EGM96 geoid. The Soviet survey map used the BK77 geoid.

*Table 2. Satellite-measurement-based Digital Elevation Model elevations and ground survey elevations of Alpomish and Peak 4643.*

| Method | Alpomish | Peak 4643 | Error |
|---|---|---|---|
| Google Terrain | 4600-4640 | 4600-4640 | ? |
| Google Earth | 4512 | 4510 | ? |
| Opentopo DEM | 4662 | 4643 | ? |
| Soviet Survey | ? | 4643.3 | ? |

Alpomish is the only other peak in Uzbekistan with a height within error bounds of peak 4643 based on the DEMs, thus this was the only other possible contender for the country highpoint. Both peaks are sharp enough that the error bounds are likely much higher than +/-16m for satellite-based DEMS.

Alpomish was not directly surveyed on the 1980 Soviet topographic map and there were in fact no previous documented ascents of the peak (Gilbertson 2024). Thus, based on all existing data, it was unknown which peak was the country highpoint.

## 1.3 Gambia

Gambia is a low-elevation country in West Africa that had not previously been surveyed carefully enough to definitively determine the location and elevation of the highest point. The country is located approximately +/-10km north and south of the Gambia river and is completely surrounded by Senegal except for a small border with the Atlantic Ocean.

Google Earth DEM data (Google Earth 2023) show the highest four locations in Gambia are all with elevations 61-65m (EGM96 geoid). These are all officially unnamed, but they will be referred to based on the districts they are located in.

They are Sare Bala Hill, Sare Doulde Hill, Sare Bissou Hill, and Sare Firasu Hill (Figure 4A, 4B). DEM data from topographic-maps.com (Yamazaki 2017) and floodmap.net (Floodmap 2021) agree with the Google Earth DEM data that these are the highest locations in the country. These peaks are within error bounds of the same elevation, thus the data is not sufficient to tell which of the four candidates are highest (Table 3).

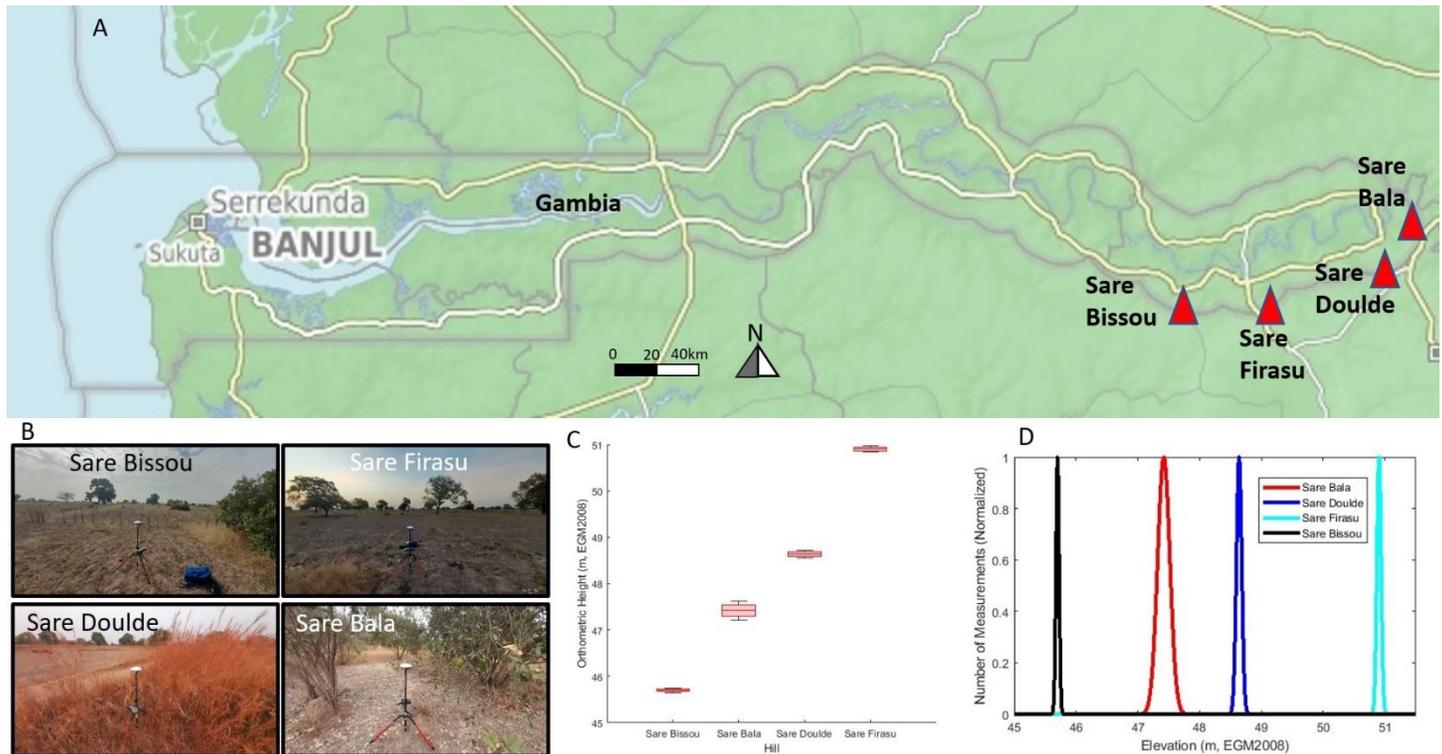

*Figure 4. (A) Map showing the locations of Sare Bissou Hill, Sare Firasu Hill, Sare Doulde Hill, and Sare Bala Hill, the four highest locations in Gambia. Basemap from OpenTopography (OpenTopography 2025). (B) Pictures of each location with dGPS set up. (C) Orthometric height of each candidate highpoint of Gambia. Box edges represent one sigma error bounds and whiskers represent two-sigma error bounds (95% confidence intervals). (D) Bell curves using measured means and standard deviations.*

Previous ground surveys conducted in 1966 (Joint 1966) and 1976 (Gambia 1976) reported the highest surveyed location at 53m (vertical datum Banjul PWD BM4) at Sare Bala Hill (location 13.391589N, -13.799529E), though error bounds were not reported. The other three highpoint candidates were not measured for these surveys, though.

Three other ground surveys have been conducted in Gambia, a 1981 Soviet survey (Soviet D28-22 1981) and two US military surveys (US ND28-11 1974, US I02-b 1994). These surveys produced topographic maps with 10m contour lines, so were also not accurate enough to distinguish between the top four candidates.

Additional uncertainty exists in the exact location of the Gambia-Senegal border, and this affects one of the highpoint candidates. The border in the region around Sare Firasu Hill is shown differently according to different sources. Google maps (Google 2021), and the 1981 soviet survey (Soviet D28-22 1981) show Sare Firasu Hill within Gambia, while the 1976 British map (Gambia 1976) show it in Senegal.

If the Google Maps/1981 Soviet survey border is correct, Sare Firasu Hill is on the Gambia-Senegal border and Google Earth DEM measures it at 65m. This is the highest candidate based on Google Earth. But if the 1976 British survey border is correct, then Sare Firasu Hill is not in Gambia and not actually a highpoint candidate.

*Table 3. DEM elevations (m) of the four candidate highpoints (EGM96 geoid).*

| Peak | Google Earth | Topographic-map | Floodmap | Gaia |
|---|---|---|---|---|
| Sare Bala Hill | 61 | 57 | 63 | 63 |
| Sare Doulde Hill | 61 | 57 | 64 | 61 |
| Sare Bissou Hill | 61 | 55 | 68 | 66 |

| Sare Firasu Hill | 65 | 61 | 68 | 67 |

According to "African Boundaries – A Legal and Diplomatic Encyclopaedia" (Brownlie 1979), written in 1979, for the relevant area of eastern Gambia the border is defined as 10km from the south bank of the Gambia river, subject to adjustments. Sare Firasu Hill is 11km from the south edge of the Gambia River, so as long as there were no adjustments after 1979 then it would be completely in Senegal.

The 1981 Soviet survey was conducted two years after the Encyclopaedia border definition and five years after the British survey, so it was possible there was a border adjustment in that time.

### 1.4  Guinea-Bissau

Guinea-Bissau has not previously been surveyed carefully enough to definitively determine which point is the highest in the country. Satellite-based DEMs (Google Earth 2021) indicate two candidate locations, Mt Ronde (location 11.683069N, -13.892175E) and an unnamed hill (location 11.739242N, -13.707989E) near the village of Venu Leidi, with elevations within measurement error bounds of each other (Figure 5A). The unnamed hill will be referred to as Venu Leidi Hill.

Existing elevation measurements for these two locations are shown in Table 4 from Google Earth DEM (Google Earth 2021) a 1981 Soviet ground survey (Soviet C28-05 1981), a 1974 US Joint operation ground survey (US NC28-03 1974), a topographic-map.com DEM (Yamazaki 2017), floodmap.net DEM (Floodmap 2021) and Gaia DEM (Gaia 2021).

The 1981 and 1974 ground surveys each found Venu Leidi Hill taller, but error bounds were not given for measurements. Satellite-based DEMs have found both peaks within error bounds of the same elevation. Existing measurements were not conclusive in determining which peak was taller.

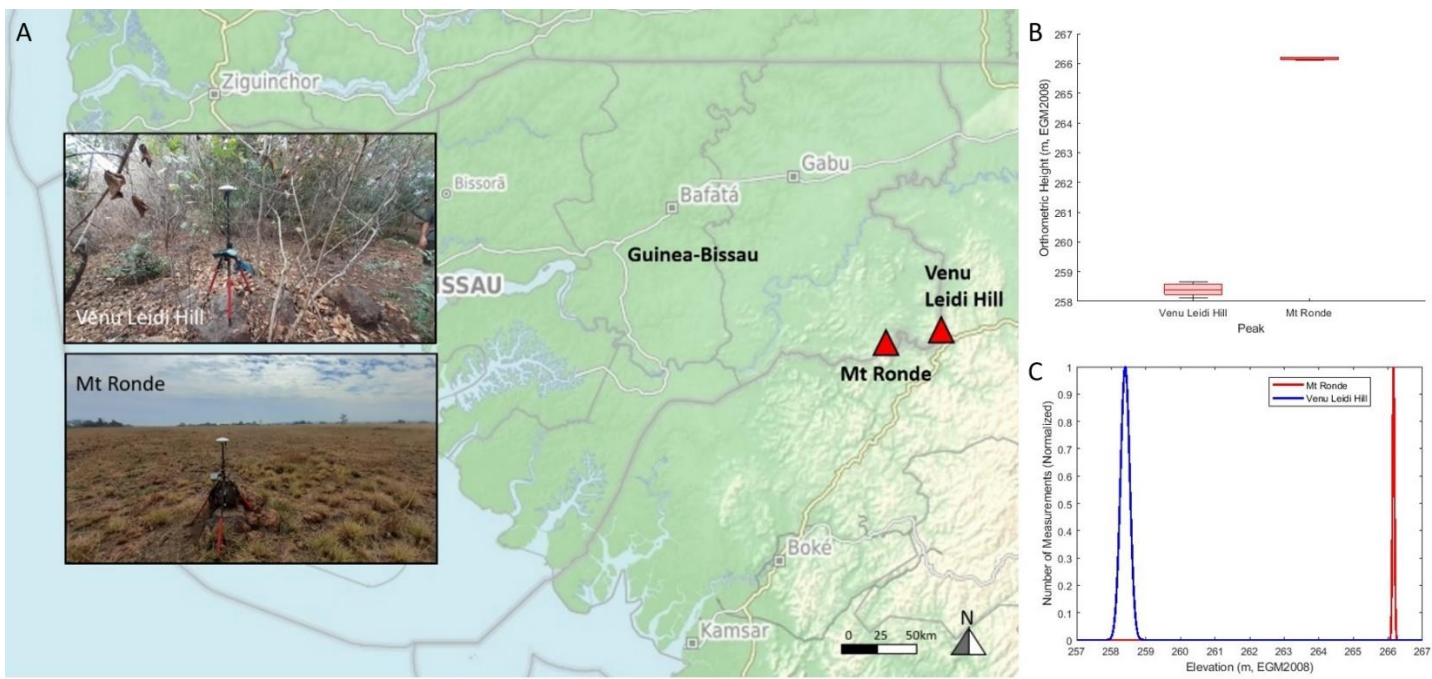

*Figure 5. (A) Map showing locations of Mt Ronde and Venu Leidi Hill, the tallest peaks in Guinea-Bissau. Basemap from OpenTopography (OpenTopography 2025). (B) Measured orthometric height of each peak. Box edges represent one-sigma error bounds and whiskers represent two-sigma errors. (C) Normalized bell curves for each measurement assuming normal distribution.*

*Table 4. Existing measurements of elevations of Mt Ronde and Venu Leidi Hill. Satellite-based DEM elevations given using EGM96 geoid. Triangulation-based measurements use unspecified geoid.*

| Survey | | | Mountain | | |
| Name | Method | Year | Mt Ronde | Venu Leidi Hill | Error (m) |
| --- | --- | --- | --- | --- | --- |
| US Joint | Triangulation | 1974 | 250 | 269 | ? |
| Soviet | Triangulation | 1981 | 262 | 292 | ? |
| Google Earth | Satellite, DEM | 2000 | 272 | 263 | +/-16 |
| ArcGIS | Satellite, DEM | 2000 | 285 | 278 | +/-16 |
| Topographic-map | Satellite, DEM | 2000 | 273 | 264 | +/-16 |
| Floodmap | Satellite, DEM | 2000 | 275 | 258 | +/-16 |
| Gaia | Satellite, DEM | 2000 | 268 | 265 | +/-16 |

## 1.5 Togo

Togo had not previously been surveyed accurately enough to determine which point is the highest in the country. Satellite-based measurements indicate the two highest candidates are Mt Agou (location: 6.872803N, 0.749182E) and Mt Atilakoutse (location: 7.329054N, 0.709001E) (Figure 6A). These are within the error bounds of the same elevations.

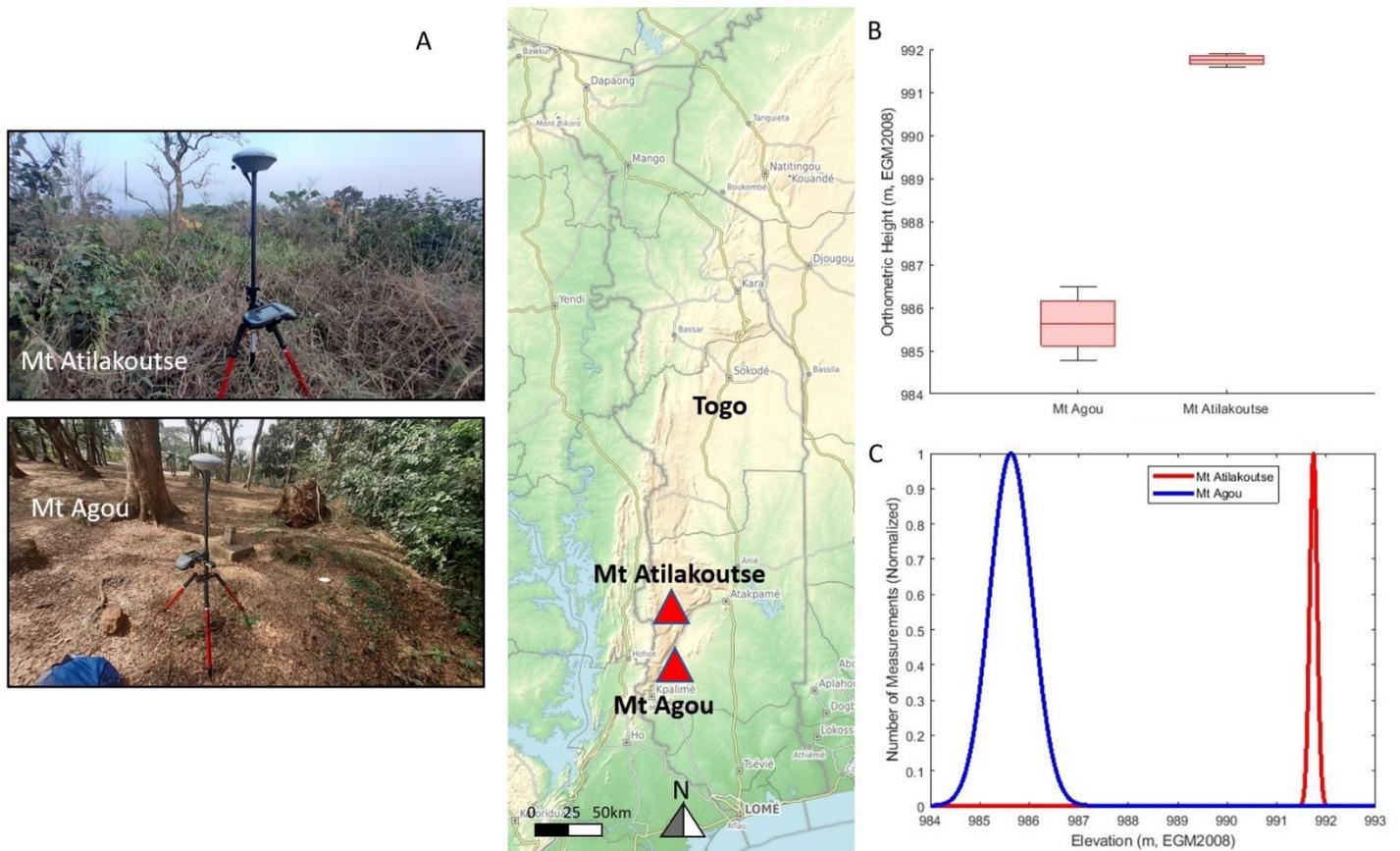

*Figure 6. (A) Map showing locations of Mt Atilakoutse and Mt Agou, the two highest peaks in Togo. Basemap from OpenTopography (OpenTopography 2025). Pictures of each peak inset. (B) Measured orthometric height of Mt Agou and Mt Atilakoutse. Box is centered about mean with box edges extending to one-sigma errors and whiskers extending to two-sigma errors (95% confidence interval). (C) Normalized bell curves centered about mean measurement for each peak, and measured standard deviations used.*

Triangulation-based ground surveys were conducted in 1981 (Soviet B31-07 1981), 1974 (US NB31-05 1974) and 1994 (US K01-D 1994), which all measured Mt Agou taller, but no error bounds were given for measurements. Google Earth DEM (Google Earth 2021), topographic-map.com DEM (Yamazaki 2017), floodmap.net DEM (Floodmap 2021), ArcGIS

DEM (ArcGIS 2021), and Gaia DEM(Gaia 2021) elevations are given in Table 5. These all measured Mt Atilakoutse of equal or higher elevation, though elevations of both peaks were within error bounds of each other (Smith 2003).

Based on existing measurements, it was not known with certainty which peak was the country highpoint.

Table 5. Elevations of Mt Agou and Mt Atilakoutse with error bounds when known. Satellite-based DEM elevations given using EGM96 geoid. Triangulation-based measurements use unspecified geoid.

| Survey | | | Mountain | | |
| --- | --- | --- | --- | --- | --- |
| Name | Method | Year | Mt Agou | Mt Atilakoutse | Error (m) |
| US Joint | Triangulation | 1974 | 986 | 961 | ? |
| Soviet | Triangulation | 1981 | 986 | 950 | ? |
| US Military | Triangulation | 1994 | 986 | 947 | ? |
| Google Earth | Satellite, DEM | 2000 | 986 | 989 | +/-16 |
| ArcGIS | Satellite, DEM | 2000 | 1002 | 1002 | +/-16 |
| Topographic-map | Satellite, DEM | 2000 | 983 | 984 | +/-16 |
| Floodmap | Satellite, DEM | 2000 | 986 | 993 | +/-16 |
| Gaia | Satellite, DEM | 2000 | 989 | 993 | +/-16 |

## 2 Methodology:

In general, for each country, the highest candidate peaks with elevations within the error bounds of satellite-based DEMs were identified. Ground measurements were taken on each candidate to determine the highest peak.

### 2.1 Saudi Arabia

Absolute elevation measurements were conducted using a Trimble Geo 7x differential GPS unit, which is capable of sub-meter vertical accuracy.

On August 17, 2018, three 10-minute measurements were taken on the summit of Jabal Sawda, and on August 18, three 10-minute measurements were taken on the summit of Jabal Ferwa (Figure 2A). On each summit the dGPS was placed on the highest rock.

Data was processed using measurements from base stations in Kuwait and Addis Ababa, Ethiopia, the closest available stations accessible via Trimble Pathfinder Office software.

One of the main factors that contributes to degradation of the accuracy of GPS measurements is distortion of the satellite signals caused by the local ionospheric conditions; i.e., the flow of charged particles high in the atmosphere, in between the GPS satellite and earth. The base station is able to calculate the ionospheric-induced distortion at its location, which can be used to correct GPS measurements taken nearby.

For each peak, Jabal Ferwa and Jabal Sawda, three sets of measurements were taken, referred to as Ferwa1, Ferwa2, Ferwa3, for Jabal Ferwa; and Sawda1, Sawda2, and Sawda3 for Jabal Sawda (Figures 2B, 2C). Two base stations were used for post-processing: 'kuwt' (UNAVCO) and 'adis' (IGS). All data were post-processed using Trimble Pathfinder Office (PFO). For 'kuwt1,' PFO automatically retrieved base station data; for 'kuwt2' and 'adis,' base station data were downloaded separately and imported into PFO.

### 2.2 Uzbekistan

On August 22, 2023, a 10 arcminute 5x Sokkia Abney level and a 10 arcminute 1x Sokkia Abney level were used to take angular inclination measurements from a tarn (location: 38.89117N, 68.19968E) with a known elevation from the 1980 topographic map to the summit of Alpomish (Figure 3B). Using the known coordinates of the summit and tarn, the distance $D$ from the tarn to the summit was calculated. Basic trigonometry was used to calculate the relative height $H$ of Alpomish above the tarn, using equation

$$H = D \tan \varphi$$

where $\varphi$ is the inclination angle. The absolute elevation, $Z_A$, of Alpomish was calculated by adding the absolute height, $Z_{tarn}$, of the tarn to the relative height of Alpomish above the tarn,

$$Z_A = Z_{tarn} + H.$$

This resulted in two absolute elevation measurements of Alpomish.

On August 23, 2023, both Abney levels and a Promark 220 differential GPS unit with Ashtech antenna were brought to the summit of Alpomish. This was the first ascent of Alpomish (Gilbertson 2024) and involved a multipitch technical rock climb.

Angular declinations were measured from the summit of Alpomish down to the summit of Peak 4643 using each Abney level (Figure 3C). The known coordinates of each summit were used to calculate the horizontal distance $d$ between peaks. Pictures of each peak are shown in Figure 3A, with the view of Alpomish from the tarn (left) and the view of Peak 4643 from the summit of Alpomish (right).

Basic trigonometry was used to calculate the relative height $\Delta$ of Alpomish above Peak 4643 using

$$\Delta = d \tan \theta . \tag{1}$$

The absolute height $Z_A$ of Alpomish was then calculated by adding the known height of peak 4643 to the measured relative height $\Delta$. This gave two more absolute height measurements for Alpomish.

The Promark 220 differential GPS unit was set up on the summit of Alpomish, but it failed to acquire satellites. Thus, no usable data was recorded. One handheld GPS unit (Garmin 62s) was able to acquire satellites and record absolute summit elevation data. This gave one more absolute height measurement for Alpomish.

On August 25 both Abney levels were brought to the summit of Peak 4643. This was the fifth documented ascent of this peak (Gilbertson 2024). Angular inclination measurements were taken from Peak 4643 to Alpomish using both Abney levels. The relative height of Alpomish above Peak 4643 was calculated using Eq (1). The absolute height $Z_A$ of Alpomish was then calculated by adding the known height of peak 4643 to the measured relative height. This gave two more absolute height measurements for Alpomish.

The Garmin 62s was used to measure the absolute height of Peak 4643.

### 2.3 Gambia

A Trimble GeoXR differential GPS unit with Zephyr 2 antenna capable of sub-meter vertical accuracy was brought to each of the four highpoint candidates and a one-hour measurement taken using a 1.0m antenna rod. The coordinates of the highpoint locations were taken from the Google Earth DEM data.

Sare Bala Hill and Sare Doulde hill were surveyed on Dec 17, 2021. Sare Firasu Hill and Sare Bissou Hill were surveyed on December 18, 2021. According to locals in Foss Bojang village, the nearest village to Sare Firasu Hill, the village and hill are both completely within Gambia. This is consistent with the border as drawn on the 1981 Soviet topographic map and on Google Maps, which were the most recent sources for the border location.

### 2.4 Guinea-Bissau

On Dec 26, 2021, a Trimble GeoXR differential GPS unit with Zephyr 2 antenna and 1.0m antenna rod was brought to the summit of Venu Leidi Hill and data collected for one hour. On Dec 27, 2021, the unit was brought to the summit of Mt Ronde and data collected for one hour.

## 2.5 Togo

On December 13, 2021, a Trimble GeoXR differential GPS unit with Zephyr 2 antenna and 1.0m antenna rod was brought to the summit of Mt Agou. Data was logged for 30 minutes (the planned one-hour measurement had to be cut short). The same day, the unit was brought to the summit of Mt Atilakoutse and data logged for one hour.

# 3   Results:

Elevations will be reported as orthometric height using the EGM2008 geoid (Pavlis et al. 2012) for Saudi Arabia, Gambia, Guinea-Bissau, and Togo. This geoid is used because it has improved accuracy compared to the EGM96 geoid. For Uzbekistan, results will be reported in the same vertical datum as the 1980 Soviet ground survey, because measurements were taken relative to elevations from that survey.

## 3.1   Saudi Arabia

Results from all measurements processed with the nearest base stations, kuwt1, kuwt2, and adis are shown in a box plot in Figure 2B. Rectangles are centered at mean height with rectangle edges at +/- one sigma and whisker edges at +/- two sigma (95% confidence interval). Rectangle color refers to base station used for post processing.

The most data points were taken in Ferwa1 (581) for Jabal Ferwa and in Sawda1 (835) for Jabal Sawda. Data processed with base station kuwt1 contained the lowest errors for each peak. Figure 2C shows a normalized histogram of elevation data from Ferwa1 and Sawda1 processed with kuwt1 base station.

This results in an elevation for Jabal Ferwa of 3,001.8m +/-1.4m, and an elevation for Jabal Sawda of 2,998.7m +/-1.2m (95% confidence interval). The two-sigma error bounds are non-overlapping, meaning Jabal Ferwa is taller than Jabal Sawda with greater than 95% confidence.

## 3.2   Uzbekistan

Both Abney level measurements from the tarn measured 19 degrees 50 minutes at a horizontal distance of 1.9km. This means a relative height of 685m was measured. Assuming an error of +/-5 arcminutes (given the 10-arcminute resolution of the Abney level), this means the vertical error was +/-3m. Adding the known elevation of the tarn, 3982m, to the measured relative height gave an absolute height of 4667m +/- 3m.

From the summit of Alpomish, both Abney level measurements were between 10-20 min declination down to Peak 4643 (the Abney level Vernier scale is in 10-min increments). The value 15 min +/-5 min will be used for calculations. The horizontal distance between the two peaks was calculated from the known coordinates of each peak. This distance was 6.3 km. This resulted in a relative height of Alpomish 25m +/-8m above Peak 4643. Adding the relative height measured to the absolute height of Peak 4643 resulted in absolute height measurements of Alpomish of 4668m +/-8m.

From the summit of Peak 4643, both Abney level measurements were 10-20min inclination up to Alpomish. The value 15 min +/-5 min will be used for calculations. These were consistent with the declination measurements from Alpomish down to Peak 4643. This means the measurements also resulted in absolute height measurements of Alpomish of 4668m +/-8m.

The horizontal distance between Alpomish and Peak 4643 is far enough that earth curvature corrections might be significant. At this distance the earth curvature correction $h$ is 3.1m in vertical height, using equation

$$h = \frac{d^2}{2R}$$

where $d$ is the horizontal distance of the measurement and R is the radius of the earth. This result will be added to the previous error bounds. Thus, the error bounds of +/-11m will be reported, rounded to nearest meter.

The Garmin 62S measured Alpomish 4m taller than Peak 4643. Error bounds were not given for this measurement. In general, handheld GPS units can have vertical errors up to +/-23m (Rodriguez et al. 2016), but they are not known for certain.

Table 6 gives all measurements and errors for each peak. Garmin 62s measurements are given as the sum of the relative height measured plus the absolute height of peak 4643, to ensure a consistent vertical datum is used. Measurements are rounded to the nearest meter.

*Table 6: Ground measurement elevations of Alpomish and Peak 4643. Note – elevation of Peak 4643 used from 1980 Soviet survey.*

| Method | Alpomish | Error (+/-m) | Peak 4643 |
|---|---|---|---|
| 1x Abney from Alpomish | 4668 | 11 | 4643 |
| 5x Abney from Alpomish | 4668 | 11 | 4643 |
| 1x Abney from 4643 | 4668 | 11 | 4643 |
| 5x Abney from 4643 | 4668 | 11 | 4643 |
| 1x Abney from tarn | 4667 | 3 | - |
| 5x Abney from tarn | 4667 | 3 | - |
| Garmin 62s | 4647 | ? | 4643 |

The final absolute height of Alpomish was calculated by taking an average of the six Abney level measurements. Results are rounded to the nearest meter. The error bounds are reported as the highest error bounds, +/-11m. The Garmin 62s measurements are not used in the average, because the error bounds are not known for those measurements. The Promark 220 did not acquire satellites, so no data was measured.

The final absolute height of Alpomish is thus 4668m +/-11m. This is in the BK77 geoid, the geoid used by the 1980 survey. The elevation of Peak 4643 is assumed to be that measured by the 1980 survey, 4643.3m. This means, rounding results to the nearest meter, Alpomish is 25m +/-11m taller than Peak 4643.

In July 2025 author EG returned to Alpomish with a Trimble DA2 differential GPS unit, which properly acquired satellites on the summit. The measured absolute height was 4651.4m +/-0.1m (95% confidence interval, EGM2008 geoid). It is not currently possible to convert between EGM2008 geoid and BK77 geoid in Uzbekistan, and this is a subject of ongoing work within Uzbekistan (Erkin 2017). The current standard geoid used in Uzbekistan is BK77, so the final reported height will be the height using the BK77 geoid, 4668m. Peak 4643 was not re-measured, so the relative height measurements from 2023 still hold, and Alpomish is still the national highpoint.

### 3.3 Gambia

Data was processed by Compass Data engineers using PPP (Precise Point Positioning) to give ellipsoidal heights and 95% confidence interval errors. The ellipsoidal elevations were converted to orthometric elevation using EGM2008 geoid (Table 7).

*Table 7. Measured elevations and two-sigma errors or top four highest locations in Gambia.*

| Peak | Elevation (m) | Two-Sigma Error (+/-m) | Latitude | Longitude |
|---|---|---|---|---|
| Sare Bissou Hill | 45.702 | 0.05 | 13.235365N | -14.356052E |
| Sare Bala Hill | 47.421 | 0.198 | 13.391589N | -13.799529E |
| Sare Doulde Hill | 48.64 | 0.085 | 13.330719N | -13.846185E |
| Sare Firasu Hill | 50.905 | 0.067 | 13.221566N | -14.159756E |

The elevation data is shown in the box plot (Figure 4C) and a normalized histogram (Figure 4D). Sare Firasu Hill has the highest measured elevation at 50.905m +/-0.067m (95% confidence interval). The error bounds of this measurement do not overlap with error bounds from any other measurements.

### 3.4 Guinea-Bissau

Results were processed by Compass Data engineers using PPP processing to give ellipsoidal heights and 95% confidence interval errors. The ellipsoidal heights were then converted to orthometric height using EGM2008 geoid (Table 8). Mt Ronde was measured at 266.161m +/-0.057m and Venu Leidi Hill was measured at 258.4m +/-0.267m.

*Table 8. Measured elevations and two-sigma errors for Mt Ronde and Venu Leidi Hill.*

| Peak | Elevation (m) | Error (Two-sigma, +/-m) | Latitude | Longitude |
|---|---|---|---|---|
| Mt Ronde | 266.161 | 0.057 | 11.683069N | -13.892175E |
| Venu Leidi Hill | 258.4 | 0.267 | 11.739242N | -13.707989E |

The elevation data is shown in the box plot (Figure 5B) and a normalized histogram (Figure 5C). Mt Ronde was measured higher than Venu Leidi Hill, with non-overlapping error bounds.

### 3.5 Togo

Data was processed by Compass Data engineers using PPP processing to give ellipsoidal heights and 95% confidence interval errors. The ellipsoidal heights were then converted to orthometric height using EGM2008 geoid. Mt Atilakoutse was measured at 991.743m +/- 0.15m and Mt Agou was measured at 985.628m +/-0.85m. (Table 9).

*Table 9. Measured elevations and two-sigma error bounds of Mt Agou and Mt Atilakoutse.*

| Peak | Elevation (m) | Error (two-sigma, +/-m) | Latitude | Longitude |
|---|---|---|---|---|
| Mt Atilakoutse | 991.743 | 0.15 | 7.329054N | 0.709001E |
| Mt Agou | 985.628 | 0.85 | 6.872803N | 0.749182E |

The elevation data is shown in the box plot (Figure 6B) and a normalized histogram (Figure 6C). Mt Atilakoutse was measured higher than Mt Agou, with non-overlapping error bounds.

## 4 Discussion

In this section, survey results are summarized for each of the five countries.

### 4.1 Saudi Arabia

Based on measurements Ferwa 1 and Sawda 1 using kuwt1 base station data, Jabal Ferwa is 3.1m taller than Jabal Sawda. The two-sigma error bounds for the measurements are non-overlapping, meaning Jabal Ferwa is taller than Jabal

Sawda with greater than 95% confidence. These results have been recognized by the Saudi Climbing Federation (Anas 2018, Ferwa 2018).

Therefore, Jabal Ferwa is the country highpoint of Saudi Arabia.

## 4.2   Uzbekistan

All seven ground measurements are consistent that Alpomish is taller than Peak 4643. This is consistent with satellite-based DEMs, which also measure Alpomish either taller than Peak 4643 or both peaks are within error bounds of each other.

Abney level elevation measurements are all consistent with the final height of 4668m +/-11m.  Because the Abney level measurements were taken relative to the 1980 surveyed elevations, the 4668m elevation is thus in the same vertical datum as the 1980 measurement, BK77.

Therefore, Alpomish is the country highpoint of Uzbekistan.

## 4.3   Gambia

Sare Firasu hill was measured to have the highest elevation of all candidates at 50.905m +/-0.067m (95% confidence interval). This location was confirmed by locals to be within Gambia. The next highest candidate, Sare Doulde Hill, was more than 2m lower. The error bounds do not overlap on these measurements, meaning Sare Firsau Hill is the highest with over 95% confidence. This result is consistent with the highest location as identified from all satellite-based DEMs.

Therefore, Sare Firasu Hill is the country highpoint of Gambia.

## 4.4   Guinea-Bissau

Mt Ronde was measured higher than Venu Leidi hill with non-overlapping two-sigma error bounds. This means Mt Ronde at 266.161m +/- 0.057m is the highest peak in Guinea-Bissau with greater than 95% confidence.

Therefore, Mt Ronde is the country highpoint of Guinea-Bissau.

## 4.5   Togo

Mt Atilakoutse was measured higher than Mt Agou with non-overlapping two-sigma error bounds. This means Mt Atilakoutse at 991.743m +/-0.15m is higher than Mt Agou with greater than 95% confidence.

Therefore, Mt Atilakoutse is the country highpoint of Togo.

The final results are summarized in Table 10.

*Table 10. Final results of location and elevation of national highpoint for each country surveyed.*

| Country | Highpoint | Location | Elevation (m) |
|---|---|---|---|
| Saudi Arabia | Jabal Ferwa | 18.266717N, 42.368264E | 3001.8 |
| Uzbekistan | Alpomish | 38.891634N, 68.176775E | 4668 |
| Gambia | Sare Firasu Hill | 13.221566N, -14.159756E | 50.9 |
| Guinea-Bissau | Mt Ronde | 11.683069N, -13.892175E | 266.2 |
| Togo | Mt Atilakoutse | 7.329054N, 0.709001E | 991.7 |

# 5   Conclusion

Ground surveys have determined the country highpoints of Saudi Arabia, Uzbekistan, Gambia, Guinea-Bissau, and Togo. These national highpoints are Saudi Arabia – Jabal Ferwa (3001.8m, location 18.266717N, 42.368264E), Uzbekistan – Alpomish (4668m, location 38.891634N, 68.176775E), Gambia – Sare Firasu Hill (50.9m, location 13.221566N, -14.159756E), Guinea-Bissau – Mt Ronde (266.2m, location 11.683069N, -13.892175E), and Togo – Mt Atilakoutse

(991.7m, location 7.329054N, 0.709001E). Knowledge of these peaks offers a valuable lens for understanding the physical geography of these countries. While human factors determine national borders, physical features of a national highpoint shape the country's natural resource availability, environmental systems, and tourism patterns. Accurate elevations are important for watershed management and climate modeling, as well as a topic of national pride in each country.

# 6 Acknowledgements

The authors would like to thank Compass Data engineers for providing equipment and processing data for measurements in Gambia, Togo, and Guinea-Bissau. Waypoint Technology provided equipment and assisted with data processing for Saudi Arabia. Seattle University provided equipment for Uzbekistan. Jake O helped process data from Saudi Arabia. Serge M, Kahler K, Andreas F, and Majed A assisted in data collection.